\documentclass[11pt,a4paper]{article}
\usepackage{jheppub}
\usepackage{amsmath,amssymb,bm}
\usepackage{gensymb}
\usepackage{nicefrac} 
\usepackage{epsfig}
\usepackage{graphicx}
\usepackage{slashed}
\usepackage{epsfig} \usepackage{graphicx} \usepackage{color}
\usepackage{mathrsfs} \usepackage{amssymb} \usepackage{amsmath} \usepackage{url}
\usepackage[mathscr]{eucal}

\usepackage{xspace} 

\title{A different approach to electroweak interactions without using any $SU(2)_L$ doublet.}

\author[a]{Anirban Karan}

\affiliation[a]{The Institute of Mathematical Sciences, HBNI,\\ Taramani, Chennai 600113, India.}

\emailAdd {kanirban@imsc.res.in}

\abstract{We know that demanding $SU(2)_L\times U(1)_Y$ gauge symmetry of Lagrangian is a ``sufficient" condition to describe electroweak interactions; however, in this paper, we have tried to find whether it's a ``necessary" condition or not. We have used a different approach to describe electroweak interactions without using any $SU(2)_L$ doublet or $SU(2)_L$ generator and incorporate some new physics aspects into theory. In this method, we only need  local gauge invariance of Lagrangian under a ``generalised $U(1)_Q$ gauge transformation", where $Q$ is electric charge (not to be confused with usual $U(1)$ transformation). Under this gauge transformation, all charged fields including  charged gauge bosons transform as $H_{(\mu)}\rightarrow H_{(\mu)}e^{iq_h\theta}$ and all  neutral gauge bosons transforms as $V_\mu\rightarrow V_\mu+\Lambda \partial_\mu \theta$. Nevertheless, $SU(2)_L$ symmetry can be restored by imposing constraints on some parameters and thus  standard model can be considered as a special case of this model. Hence, it turns out that $SU(2)_L\times U(1)_Y$ gauge symmetry is a ``sufficient" but ``not necessary" condition for electroweak interactions.}

\allowdisplaybreaks

\begin{document}
\maketitle
\flushbottom

\section{Introduction}
\label{intr}

Electroweak interaction is one of widely studied topics in high energy physics, both theoretically and experimentally. Weak interactions are described using non abelian gauge theory\cite{yang mills} whereas electromagnetic interactions are explained by abelian $U(1)_Q$ gauge theory. Standard model (SM) gives good theoretical explanation of electroweak interactions taking $SU(2)_L\times U(1)_Y$ gauge symmetry of Lagrangian \cite{glashow,salam,Wienberg,barger,peskin,cheng,griffiths}. It implies that demanding $SU(2)_L\times U(1)_Y$ gauge symmetry of Lagrangian is a ``sufficient" condition to describe electroweak interactions. However, recent experiments, showing deviations of data from SM predictions in several cases \cite{k*1,k*2,k*3,d*1,d*2,d*3,neutrino1,neutrino2,neutrino3,neutrino4,fifth1}, leads us to ask whether  $SU(2)_L\times U(1)_Y$ gauge symmetry is a ``necessary" condition or not for describing electroweak interactions. In this paper, we have tried to figure out the answer to that query.

 Usually, some extra symmetries are imposed on the Lagrangian to explain deviations of data theoretically, keeping the $SU(2)_L\times U(1)_Y$ gauge symmetry intact. Our approach differs from the existing methods largely. Still, it can produce same kind of theoretical results as SM and in addition, it  can also incorporate different physics aspects beyond Standard Model (BSM) like new electroweak gauge bosons  \cite{fifth1,fifth2,new-bos1,new-bos2,new-bos3,new-bos4,fcnc1,fcnc2,fcnc3,fcnc4}, non-universality and non-unitary mixing matrices for quarks or leptons  \cite{non-uni1,non-uni2,non-uni3,non-uni4,non-uni5,non-uni6}, flavour changing neutral current (FCNC) at tree level \cite{fcnc1,fcnc2,fcnc3,fcnc4}  etc. In this approach, we start with the assumption that $SU(2)_L$ symmetry is not an exact symmetry in the energy scale, we generally deal in experiments; rather it seems to be exact as BSM effects are tiny in most of the cases. So, it is possible that there exists some different symmetry in the Lagrangian at higher energy scale and in low energy approximation that symmetry resembles $SU(2)_L$. Thus we start our approach with a different symmetry and let us call it ``generalised $U(1)_Q$ gauge symmetry." We take all charged fields including fundamental charged vector bosons to be transformed as $H_{(\mu)}\rightarrow H_{(\mu)}e^{iq_h\theta}$ and all fundamental neutral vector bosons to be transformed as $V_\mu\rightarrow V_\mu+\Lambda \partial_\mu \theta$ under  ``generalised $U(1)_Q$ gauge transformation" ($Q$ denotes electric charge,  $q_h$ is the electric charge of particle $H$ and $\Lambda$ is some constant). Then, we demand that the Lagrangian must be invariant under this gauge transformation. How this claim eventually lead to electroweak interactions is the main objective of this paper. This technique can be thought as an extension of Quantum Electrodynamics (QED). We emphasise that any kind of $SU(2)_L$ doublets or $SU(2)_L$ generators are not used at all in our method. We only look for the effects of ``generalised $U(1)_Q$ gauge symmetry" on the Lagrangian in this model. However, constraints on some parameters of this model can lead to usual $SU(2)_L$ symmetry and thus SM can be considered as a special case of this model.
 
It took several decades to build a sound model (both theoretically and experimentally)
for electroweak interactions (Standard Model) through numerous ingenious ideas; yet we get experimental evidences for various New Physics effects. So, one should not expect that a single paper will explain all theoretical and experimental aspects of electroweak interactions abandoning SM completely. Instead of imagining this model as an alternative to SM, one should consider it as the first step to look into electroweak interactions from a different perspective.

\section{Interactions of gauge bosons}
\label{igb}

Let us first assume that there exist two complex fundamental vector boson fields ($A_+^{\mu}$ and $A_-^{\mu}$) which are charge conjugate to each other. As we are not trying to formulate any effective field theory, we do not take $A_\pm^{\mu}$ to be any composite vector bosons (like vector mesons). In the beginning, we are not labelling them as ``gauge bosons" but we have discussed later why they should be taken as gauge bosons. On the other hand, to avoid non-renormalizability we will consider dimension (mass) four operators only throughout the analysis.

Now, we can write $A_\pm^{\mu}$ as complex linear combinations of two real vector boson fields $A_1^\mu$ and $A_2^\mu$ in the following manner
\begin{equation}
\label{eq:def12}
A_+^{\mu}=\frac{A_1^\mu-iA_2^\mu}{\sqrt{2}} \quad \textrm{and} \quad A_-^{\mu}=\frac{A_1^\mu+iA_2^\mu}{\sqrt{2}}
\end{equation}

The free field Lagrangian for these fundamental vector bosons without any kind of interaction looks like
\begin{equation}
\begin{split}
\label{eq:lag}
\mathcal{L}_{gauge}  =-\frac{1}{4}(F_{1\mu\nu}F_1^{\mu\nu}+F_{2\mu\nu}F_2^{\mu\nu})=-\frac{1}{2}(F_+)_{\mu\nu}F_-^{\mu\nu}
\end{split}
\end{equation}
where $F_{j}^{\mu\nu}=\partial^{\mu}A_{j}^\nu-\partial^{\nu}A_{j}^\mu$ with $j=+,-,1,2$.

Here we see that $\mathcal{L}_{gauge}$ is invariant under global gauge transformation $A_\pm^\mu\rightarrow A_\pm^\mu e^{\pm iq\theta}$ where $\theta$ is a space-time independent real quantity and $\pm q$ is the charge of $A_\pm^\mu$.
Now we demand this Lagrangian to be invariant under Local gauge transformation 
\begin{equation}
\label{eq:local}
A_\pm^\mu\rightarrow A_\pm^\mu e^{\pm i q\theta(x)} 
\end{equation}
Under this transformation
\begin{eqnarray}
\partial^\mu A_\pm^\nu \rightarrow (\partial^\mu A_\pm^\nu) e^{\pm iq\theta} \pm iq(\partial^\mu \theta) A_\pm ^\nu e^{\pm iq\theta},\\
F_\pm^{\mu\nu} \rightarrow  F_\pm^{\mu\nu}e^{\pm iq\theta}\pm iq e^{\pm iq\theta} [(\partial^\mu \theta) A_\pm ^\nu-(\partial^\nu \theta) A_\pm ^\mu],
\end{eqnarray}
Using these transformations on Eq.~\eqref{eq:lag}, we get 
\begin{equation}
\label{lgtr}
\mathcal{L}_{gauge}\rightarrow \mathcal{L}_{gauge}+\delta_1+\delta_2+\delta_3
\end{equation}
\begin{equation}
\label{deltas}
\begin{split}
&\delta_1=iq[\partial_\mu\theta(\partial^\mu A_+^\nu)A_{-\nu} - \partial_\nu\theta(\partial^\mu A_+^\nu)A_{-\mu}],\\&
\delta_2= -iq[\partial_\mu\theta(\partial^\mu A_-^\nu)A_{+\nu} -\partial_\nu\theta(\partial^\mu A_-^\nu)A_{+\mu} ],\\& \delta_3=-q^2[A_{+\mu} A_-^\mu \partial_\nu\theta \partial^\nu\theta -A_{+\mu} A_-^\nu \partial_\nu\theta \partial^\mu\theta]
\end{split}
\end{equation}

We see from Eq.~\eqref{lgtr} and Eq.~\eqref{deltas} that $\mathcal{L}_{gauge}$ is not invariant under this transformation and some extra terms appear in the transformed free Lagrangian. Hence, we need at least one more real ``gauge boson" field $A_3^\mu$ which will transform as  
\begin{equation}
\label{eq:a3tran}
A_3^\mu\rightarrow A_3^\mu+\lambda\partial^\mu \theta \quad \textrm{($\lambda$ is a constant)} \quad
\end{equation}
in order to cancel those extra terms in Lagrangian, because $\partial^\mu \theta$ is a vector quantity. As the extra terms in Eq.~\eqref{deltas} contain products of $A_+^\mu$, $A_-^\mu $ and $\partial^\mu \theta$, we can expect that $A_\pm^\mu$ and $A_3^\mu$ will interact with each other. So, we need to redefine our Lagrangian accordingly.

Observing Eq.~\eqref{lgtr}, Eq.~\eqref{deltas} and Eq.~\eqref{eq:a3tran}, it is obvious that we must add two interaction terms $T_1,$ $T_2$ in the Lagrangian to cancel $\delta_1$ and $\delta_2$ in the transformed Lagrangian where
\begin{equation}
\label{T12}
\begin{split}
&T_1=-\frac{iq}{\lambda}[A_{3\mu}(\partial^\mu A_+^\nu)A_{-\nu} - A_{3\nu}(\partial^\mu A_+^\nu)A_{-\mu}]\\&
T_2=\frac{iq}{\lambda}[A_{3\mu}(\partial^\mu A_-^\nu)A_{+\nu} -A_{3\nu}(\partial^\mu A_-^\nu)A_{+\mu} ]
\end{split}
\end{equation}
Thus, we define new Lagrangian $\mathcal{L'}_{gauge}$ as 
\begin{equation}
\label{lagp}
\mathcal{L'}_{gauge}=-\frac{1}{2}(F_+)_{\mu\nu}F_-^{\mu\nu}+T_1+T_2
\end{equation}
But, after applying generalized $U(1)_Q$ gauge transformation on $\mathcal{L'}_{gauge}$ we find that it's still not gauge invariant and 
\begin{equation}
\label{lagptr}
\mathcal{L'}_{gauge}\rightarrow \mathcal{L'}_{gauge}-\delta_3+\Delta_3
\end{equation}
where,
\begin{equation}
\label{Delt}
\begin{split}
\Delta_3=\frac{q^2}{\lambda}[2(A^\mu_3\partial_\mu\theta)&(A_+\cdot A_-)-(A_+^\mu\partial_\mu\theta)(A_3\cdot A_-)-(A_-^\mu\partial_\mu\theta)(A_3\cdot A_+)]
\end{split}
\end{equation}

Looking at Eq.~\eqref{deltas}, Eq.~\eqref{eq:a3tran}, Eq.~\eqref{lagptr} and Eq.~\eqref{Delt}, we find that we must add another interaction term $T_3$ in the Lagrangian to cancel all the extra terms in transformed  $\mathcal{L'}_{gauge}$, where
\begin{equation}
\label{T3}
T_3=-\frac{q^2}{\lambda^2}[(A_+\cdot A_-)A_3^2-(A_3\cdot A_-)(A_3\cdot A_+)]
\end{equation}
But, there might be some other terms which are gauge invariant by themselves (e.g. $F_{3\mu\nu}F_3^{\mu\nu}$) in the Lagrangian. We denote them as $\mathcal{L}_{inv}$. Thus we get the ``gauge invariant Lagrangian" as
\begin{equation}
\label{lagtg}
\widetilde{\mathcal{L}}_{gauge}=\mathcal{L}_{inv}-\frac{1}{2}(F_+)_{\mu\nu}F_-^{\mu\nu}+T_1+T_2+T_3
\end{equation}
 
Now, using Eq.~\eqref{eq:def12},Eq.~\eqref{T12} and Eq.~\eqref{T3}, we express $T_1,T_2$ and $T_3$ in terms of $A_1,A_2$ and $A_3$. Then,
\begin{equation}
\label{T1n}
\begin{split}
T_1&=-\frac{q}{\lambda}(\partial^\mu A^\nu_1)[A_{2\mu}A_{3\nu}-A_{3\mu}A_{2\nu}]\\&=-\frac{q}{2\lambda}F_1^{\mu\nu}[A_{2\mu}A_{3\nu}-A_{3\mu}A_{2\nu}]
\end{split}
\end{equation}
\begin{equation}
\label{T2n}
\begin{split}
T_2&=-\frac{q}{\lambda}(\partial^\mu A^\nu_2)[A_{3\mu}A_{1\nu}-A_{1\mu}A_{3\nu}]\\&=-\frac{q}{2\lambda}F_2^{\mu\nu}[A_{3\mu}A_{1\nu}-A_{1\mu}A_{3\nu}]
\end{split}
\end{equation}
\begin{equation}
\label{T3n}
\begin{split}
T_3&=-\frac{q^2}{2\lambda^2}[(A_1^2+A_2^2)A_3^2-(A_1\cdot A_3)^2-(A_2\cdot A_3)^2]\\&=-\frac{q^2}{4\lambda^2}[(A_{2\mu}A_{3\nu}-A_{3\mu}A_{2\nu})(A_{2}^\mu A_{3}^\nu-A_{3}^\mu A_{2}^\nu)\\&\quad\quad\quad+(A_{3\mu}A_{1\nu}-A_{1\mu}A_{3\nu})(A_{3}^\mu A_{1}^\nu-A_{1}^\mu A_{3}^\nu)]
\end{split}
\end{equation}
If we define new operators $\mathscr{F}_j^{\mu\nu}$ as
\begin{equation}
\label{eq:curlf}
\begin{split}
\mathscr{F}_j^{\mu\nu}=F_j^{\mu\nu}+ \frac{q}{\lambda} \epsilon_{jkl} A_k^\mu A_l^\nu
\end{split}
\end{equation}
where $(j,k,l)$ can take values $(1,2,3)$ and $\epsilon_{jkl}$ is usual Levi-Civita tensor, then using Eq.~\eqref{eq:lag}, Eq.~\eqref{lagtg}, Eq.~\eqref{T1n}, Eq.~\eqref{T2n} Eq.~\eqref{T3n} and Eq.~\eqref{eq:curlf}, we can rewrite $\widetilde{\mathcal{L}}_{gauge}$ as 
\begin{equation}
\label{lag1}
\widetilde{\mathcal{L}}_{gauge}=\mathcal{L}_{inv}-\frac{1}{4}\sum_{j=1,2}\mathscr{F}_j^{\mu\nu}\mathscr{F}_{j\mu\nu}
\end{equation}

Instead of starting with $A_1^\mu$ and $A_2^\mu$, we could have started with $A_2^\mu$ and $A_3^\mu$ and made the complex fields as linear combinations of them. Then, of course, we would have got $A_1^\mu$ as the extra ``gauge boson", required to make the Lagrangian gauge invariant. In that case, everything would go in the same way and we would end up with 
\begin{equation}
\label{lag2}
\widetilde{\mathcal{L}}_{gauge}=\mathcal{L'}_{inv}-\frac{1}{4}\sum_{j=2,3}\mathscr{F}_j^{\mu\nu}\mathscr{F}_{j\mu\nu}
\end{equation}
Comparing  Eq.~\eqref{lag1} and Eq.~\eqref{lag2} we find that
\begin{equation}
\label{inv}
\mathcal{L}_{inv}=\mathcal{L}_0-\frac{1}{4}\mathscr{F}_{3\mu\nu}\mathscr{F}_3^{\mu\nu} \textrm{  and  }\mathcal{L'}_{inv}=\mathcal{L}_0-\frac{1}{4}\mathscr{F}_{1\mu\nu}\mathscr{F}_1^{\mu\nu}
\end{equation}
\begin{equation}
\label{lagf}
\implies \widetilde{\mathcal{L}}_{gauge}=\mathcal{L}_0-\frac{1}{4}\sum_{j=1,2,3}\mathscr{F}_j^{\mu\nu}\mathscr{F}_{j\mu\nu}
\end{equation}
where $\mathcal{L}_0$ denotes the kinetic and interaction terms for other gauge bosons (like $B^\mu$ or new gauge bosons) and it must be gauge invariant.

Now, we define weak coupling constant $g$ as 
\begin{equation}
\label{eq:glam1}
g=-\frac{q}{\lambda}
\end{equation}
and thus we arrive at same Lagrangian as it should be in case of $SU(2)$ gauge symmetry where $A_\pm^\mu,A_3^\mu $ act as $W_\pm^\mu,W_3^\mu$. The main reason for using non abelian gauge theory (in SM) for electroweak interaction was to explain the interactions among gauge bosons which cannot be described by usual abelian gauge theory. However, similar kind of interactions among gauge bosons have appeared here from a completely different view point. Nevertheless, it's a coincidence that the Lagrangians for gauge fields in $SU(2)$ symmetry and generalized $U(1)_Q$ gauge symmetry are same; the Lagrangians for other interactions will be different in case of these two symmetries (as discussed in later sections). At this point one can ask why $A_3^\mu$ is identified as $W_3^\mu$, not as photon. We know from QED that under $U(1)_Q$ gauge transformation photon changes as $A^\mu\rightarrow A^\mu-\partial^\mu \theta$ when $\psi\rightarrow \psi e^{iq_f\theta}$ with the interaction term $-q_f\overline\psi\gamma^\mu\psi A_\mu$\cite{barger,peskin,cheng}. So looking at the value of $\lambda$ in Eq.~\eqref{eq:glam1} one can infer that $A_3^\mu$ is not photon. 

 In our approach, we have introduced $A_3^\mu$ to make $\mathcal{L}_{gauge}$ a gauge invariant object; so $A_3^\mu$ is a ``gauge boson". Again, we have discussed earlier how $A_1^\mu$ also can act as a ``gauge boson". Similarly, there can arise a case where $A_2^\mu$ acts as a ``gauge boson". These three cases are equivalent. So, we should take all of the three bosons on an equal footing. In this sense, there arises some kind of symmetry among $A_1^\mu$, $A_2^\mu$ and $A_3^\mu$ and all of them should be taken as gauge bosons.

\section{Charged current interactions}
\label{cci}

Now, we have to look at the interactions of these gauge bosons with fermions.  Let, there is a fermionic field $\psi$ with mass $m$. The corresponding free field Lagrangian is
\begin{equation}
\label{eq:lagferm}
\widetilde{\mathcal{L}}_{fermion}=\overline{\psi}(i\slashed{\partial}-m) \psi
\end{equation}

Taking analogy from QED, we take a general structure for any weak current to be $\overline\psi_j\gamma_\mu\psi_k$ where $j,k$ denote different fermions and search for the constraints coming from gauge invariance of Lagrangian. But at low energy scale, charged weak interactions are observed to be left handed only. This can be taken into account by taking the vertex factors for charged weak interactions to be proportional to $\gamma_\mu(1-\gamma_5)$ or charged weak currents to be $\overline\psi_{jL}\gamma_\mu\psi_{kL}$ where $\psi_L=\frac{1-\gamma_5}{2}\psi$ and $\psi_R=\frac{1+\gamma_5}{2}\psi$. In other words, the couplings of all right handed fermions with $A_\pm^\mu$  are taken to be zero in order to match the theory with observations. So we take the Lagrangian for charged current interactions as
\begin{equation}
\label{eq:lagchar1}
\widetilde{\mathcal{L}}_{ch}=-[g_{jk}\overline\psi_{jL}\gamma_\mu\psi_{kL} A_+^\mu
+g_{jk}^*\overline\psi_{kL}\gamma_\mu\psi_{jL} A_-^\mu]
\end{equation}
where $g_{jk}$ is the charged current coupling between $\psi_{jL}$ and $\psi_{kL}$. As the Lagrangian is hermition, the coupling for charged current mediated by $A_-^\mu$ is taken to be $g_{jk}^*$, complex conjugate of $g_{jk}$.

Now, according to our convention, under $U(1)_Q$ gauge transformation $\psi_j$ changes as $\psi_j\rightarrow \psi e^{iq_{j}\theta}$. So, $\widetilde{\mathcal{L}}_{ch}$ transforms as, 
\begin{equation}
\widetilde{\mathcal{L}}_{ch}\rightarrow -g_{jk}\overline\psi_{jL}\gamma_\mu\psi_{kL} A_+^\mu e^{i\Delta_q\theta}
-g_{jk}^*\overline\psi_{kL}\gamma_\mu\psi_{jL} A_-^\mu e^{-i\Delta_q\theta}
\end{equation}
where $\Delta_q=q-q_j+q_k$. Then, the condition that must hold to respect the global symmetry is
\begin{equation}
\label{eq:charge}
\Delta_q=q-q_j+q_k=0
\end{equation}

This Eq.~\eqref{eq:charge} signifies electric charge conservation at fermion-$A_\pm^\mu$ interaction vertex. In SM, there are total 24 fermions including particles and anti-particles both. Eq.~\eqref{eq:charge} indicates which two of them will interact with each other through $A_{\pm}^\mu$.

As in SM $W_\pm^\mu$ have unit charge so charged weak interaction only happens between up type quark (lepton) and down type quark (leptons) and similarly for anti-particles. But, the interacting fermions need not to be of same generation. Thus mixing of different generations comes automatically. There exists an ambiguity in lepton sector as neutrinos and anti-neutrinos both are neutral in charge. This can be solved either by demanding total lepton number conservation or by assuming neutrinos to be Majorana particle.

It is always possible to write the couplings in terms of weak coupling constant $g$. Then, charged weak interaction can be written as,
\begin{equation}
\begin{split}
\label{eq:lagchar}
\widetilde{\mathcal{L}}_{ch}=-\frac{g}{\sqrt{2}}[ V_{jk}\overline\psi_{jL}\gamma_\mu \psi_{kL} A_+^\mu
+V_{jk}^*\overline\psi_{kL}\gamma_\mu\psi_{jL} A_-^\mu]\delta_{q+q_k,q_j}
\end{split}
\end{equation}

 $V_{jk}$  in the above Eq.~\eqref{eq:lagchar} gives the strength of interaction  for two different fermions with $A_+^\mu$ in terms of $g$. At this point there is no restriction on $V_{jk}$. Now, one can define two column matrices $\Psi_u$ and $\Psi_d$ with up types quarks (or leptons) and  down type quarks (or leptons) respectively and write the charged current interaction with $A_+^\mu$ as $\overline\Psi_{uL}\gamma_\mu V \Psi_{dL} A_+^\mu$ where $V$ is a matrix made with $ V_{jk}$. Using singular value decomposition (SVD), any complex matrix can be made diagonal by multiplying two appropriate unitary matrices on both sides of it.  Then we take unitary transformations on $\Psi_u$ and $\Psi_d$ such that $V$ becomes diagonal in that basis, i.e., $\Psi_u'=U_u \Psi_u$, $\Psi_d'=U_d \Psi_d$ and $U_u V U_d^\dagger=V_{d}$ where $V_d$ is a diagonal matrix. We name this basis of fermions as ``interaction basis." Now, if we assume that the couplings for all allowed charged current interactions with $A_+^\mu$ in interaction basis are same (i.e. universality of fermions) then $V_d$ becomes proportional to identity matrix and $V$ becomes proportional to some unitary matrix. If we further assume  that the coupling in interaction basis is $g/\sqrt{2}$ for all allowed charged current interaction then $V$ becomes unitary and  we can diagonalize it by taking any one of $U_u$ and $U_d^\dagger$ as identity matrix and the remaining one as $V^\dagger$.  In SM, we assume universality of fermions inevitably as we take transformations on all doublets with same coupling ($g$). Here, we do not take any of these assumptions and leave the universality of fermions as well as the unitarity of $V$ for experimental verifications.  

\section{Neutral current interactions} 
\label{nci}

 As initially $A_1^\mu,$ $A_2^\mu$ and $A_3^\mu$ were equivalent, it is legitimate to assume that  only left handed fermions interact with $A_3^\mu$. So, we take general structure of neutral current interaction as $\overline\psi_{jL}\gamma_\mu\psi_{kL} A_3^\mu$ look at the consequences of demanding gauge invariance. From global gauge invariance, we have $q_j=q_k$. Now, we write down the whole Lagrangian as
\begin{equation}
\label{eq:lagp}
\mathcal{L}'=\widetilde{\mathcal{L}}_{gauge}+ \overline{\psi}_j(i\slashed{\partial}-m_j)\psi_j+\widetilde{\mathcal{L}}_{ch}-\widetilde{g}_{jk}\overline\psi_{jL}\gamma_\mu\psi_{kL} A_3^\mu\delta_{q_j,q_k}
\end{equation}
where $\widetilde{g}_{jk}$ is the neutral current coupling between  $\psi_{jL}$ and $\psi_{kL}$. Due to hermiticity of $\mathcal{L}'$ we need $\widetilde{g}_{kj}=\widetilde{g}_{jk}^*$. If $q_j$ is charge of $\psi_j$, under local gauge transformation given by Eq.~\eqref{eq:local} and Eq.~\eqref{eq:a3tran}, $\mathcal{L}'$ changes as
\begin{equation}
\begin{split}
\label{eq:lagtrp}
\mathcal{L}'\rightarrow \mathcal{L}'-q_j\overline\psi_{jL}\gamma_\mu\psi_{jL} \partial^\mu \theta-q_j\overline\psi_{jR}\gamma_\mu\psi_{jR} \partial^\mu \theta
 -\lambda \widetilde{g}_{jk}\overline\psi_{jL}\gamma_\mu\psi_{kL} \partial^\mu \theta\delta_{q_j,q_k} 
\end{split}
\end{equation}

From the above Eq.~\eqref{eq:lagtrp} it is evident that $\mathcal{L}'$ is not a local gauge invariant object. So,  to cancel the extra pieces in the transformed Lagrangian, we need at least one more neutral ``gauge boson" ($B_\mu$), which transforms as $B_\mu\rightarrow B_\mu+\lambda_b \partial_\mu \theta$ under the local gauge transformation. We also need to introduce suitable interaction terms for $B_\mu$ with fermions to make the Lagrangian gauge invariant and looking at Eq.~\eqref{eq:lagtrp}, we can easily write down the interaction terms to be of the form $\overline\psi_{jR}\gamma_\mu\psi_{jR}B^\mu$ and $\overline\psi_{jL}\gamma_\mu\psi_{kL}B^\mu \delta_{q_j,q_k}$. The approach for introduction of $B^\mu$ is quite similar to the approach for introduction of $A_3^\mu$; the only difference is that $A_3^\mu$ makes $\widetilde{\mathcal{L}}_{gauge}$ to be gauge invariant, whereas $B^\mu$ makes the remaining part of the Lagrangian to be gauge invariant. Now the question arises whether we need 
$\overline\psi_{jL}\gamma_\mu\psi_{jL} B^\mu$ kind of interaction at all or we can take $(q_j+\lambda \widetilde{g}_{jj})=0$. The answer is that we need this kind of interaction and $(q_j+\lambda \widetilde{g}_{jj})\neq0$. Actually, we introduced $A_3^\mu$ to make $\mathscr{L}_{gauge}$ a gauge invariant object. So, $A_3^\mu$ should not be expected to make other interactions gauge invariant. One important point to note is that $B_\mu$ should not couple to $A_\pm^\mu$ or $A_3^\mu$, otherwise the gauge invariance achieved in $\widetilde{\mathcal{L}}_{gauge}$ will be gone and we have to start from scratch again. So, we put a kinetic term $-\frac{1}{4} \mathscr{B}^{\mu\nu}\mathscr{B}_{\mu\nu}$ in  $\mathcal{L}_0$, Eq~\eqref{lagf}, where $\mathscr{B}^{\mu\nu}=\partial^{\mu}B^\nu-\partial^{\nu}B^\mu$. Then we can write  $\mathcal{L}_0=\widetilde{\mathcal{L}}_0-\frac{1}{4} \mathscr{B}^{\mu\nu}\mathscr{B}_{\mu\nu}$. Another important point is to note that
there exists left handed FCNC at tree level in this approach. The left handedness of FCNC at tree level is clear from the interactions of fermions with $A_3^\mu$ and $B^\mu$.

Let us keep aside FCNC for a while and focus on usual flavour conserving neutral currents first. So, after dropping FCNC terms and different gauge invariant terms (i.e. $\widetilde{\mathcal{L}}_{gauge}$, $\widetilde{\mathcal{L}}_{ch}$ and $m\overline{\psi}\psi$) from Eq.~\eqref{eq:lagp}, we add the flavour conserving neutral current interactions for $B^\mu$ and write 
\begin{equation}
\label{eq:lagplf}
\widetilde{\mathcal{L}}'_\tau= \overline{\psi}_\tau(i\slashed{\partial})\psi_\tau-\widetilde{g}_{f\tau}\overline\psi_\tau\gamma_\mu\psi_\tau A_3^\mu-t_{f\tau}\overline\psi_\tau\gamma_\mu\psi_\tau B^\mu
\end{equation}
where $\tau=L,R$ and $\widetilde{g}_{fR}=0$. Here we have dropped the fermion index $j$. The couplings $\widetilde{g}_{f\tau}$ and $t_{f\tau}$ depend on chirality as well as flavour of the fermions. For hermiticity of Lagrangian, $\widetilde{g}_{f\tau}$ and $t_{f\tau}$ must be real.

As we have assumed that $B^\mu$ changes as $B_\mu\rightarrow B_\mu+\lambda_b \partial_\mu \theta$
under the gauge transformation, then demanding local gauge invariance of $\widetilde{\mathcal{L}}'_\tau$ will lead to 
\begin{equation}
\label{eq:nishi1}
q_f+\widetilde{g}_{f\tau}\lambda+\lambda_b t_{f\tau}=0
\end{equation}
where $q_f$ is charge of $\psi$. Now, it is always possible to write $\widetilde{g}_{f\tau}$, $\lambda_b$ and $t_{f\tau}$ as
 \begin{equation}
 \label{eq:lamtf}
\widetilde{g}_{f\tau}=g\chi_{f\tau} \quad\textrm {and} \quad\lambda_b=-\frac{q}{g'}\quad\textrm {and} \quad t_{f\tau}=g' \xi_{f\tau}
 \end{equation} 
 where $g'$ is some constant, $\chi_{f\tau}$ and $\xi_f$ are some parameters which depend on flavour and chirality both of a fermion. Then combining the above Eq.~\eqref{eq:nishi1} with Eq.~\eqref{eq:glam1} and Eq.~\eqref{eq:lamtf} one would get
\begin{equation}
\label{eq:nishi2}
q_f=q(\chi_{f\tau}+\xi_{f\tau})
\end{equation}
So, for right handed fermions, $\chi_{fR}=0$ and $\xi_{fR}=q$ as $A_3^\mu$ couples to left handed fermions only. In case of SM, one can identify $\chi_{f\tau}$ with the value of $T_3$ (third component of weak isospin), $\xi_{f\tau}$ with $\frac{Y}{2}$ (half of weak hypercharge) and $q$ with unit charge $e$. Then the above Eq.~\eqref{eq:nishi2} indicates to Gell-Mann-Nishijima formula \cite{Nishijima,barger,griffiths}. 

Now, one can mix $A_3^\mu$ and $B^\mu$ through Wienberg angle $(\theta_W)$ to get photon ($A^\mu$) and $Z$-boson ($Z^\mu$) \cite{glashow,salam,barger,griffiths}as 
\begin{eqnarray}
\label{eq:Z}
Z^\mu&=&A_3^\mu \cos\theta_W-B^\mu \sin\theta_W\\
\label{eq:A}
A^\mu&=&A_3^\mu \sin\theta_W+B^\mu \cos\theta_W
\end{eqnarray}
Solving for $A_3^\mu$ and $B^\mu$ from Eq.~\eqref{eq:Z} and Eq.~\eqref{eq:A} one can write the interaction part of $\widetilde{\mathscr{L}}'_\tau$, in Eq.~\eqref{eq:lagplf} in terms of $A^\mu$ and $Z^\mu$ as
\begin{equation}
\label{eq:lagneu}
\begin{split}
&\widetilde{\mathcal{L}}_{nuet,\tau}=-\big[(g\chi_{f\tau} \sin\theta_W+g'\xi_{f\tau} \cos\theta_W)A^\mu\\&
\quad\quad\quad\quad+(g\chi_{f\tau} \cos\theta_W-g'\xi_{f\tau} \sin\theta_W)Z^\mu\big] \overline{\psi}_{f\tau}\gamma_\mu \psi_{f\tau}
\end{split}
\end{equation}
Identifying the coupling of photon  in the above Eq.~\eqref{eq:lagneu} to be $q_f$, one have 
\begin{equation}
\label{eq:qf}
q_f=g\chi_{f\tau} \sin\theta_W+g'\xi_{f\tau} \cos\theta_W
\end{equation}
As all particles satisfy Eq.~\eqref{eq:nishi2} and Eq.~\eqref{eq:qf} both, then one must have
\begin{equation}
\label{eq:ggp}
g=\frac{q}{\sin\theta_W}\quad\textrm{and}\quad g'=\frac{q}{\cos\theta_W}
\end{equation}
Using Eq.~\eqref{eq:lagneu} and Eq.~\eqref{eq:ggp} one can  write the  interaction of fermions with $Z$ boson to be
\begin{eqnarray}
 \widetilde{\mathcal{L}}_Z^{}=\sum_{\tau=L,R}g_z\big[\chi_{f\tau}-(q_f/q) \sin^2\theta_W\big]\overline{\psi}_{f\tau}\gamma_\mu \psi_{f\tau}\\
 \label{eq:gz}
\textrm{with}\quad g_z=\frac{q}{\sin\theta_W \cos \theta_W} \quad\quad\quad\quad
 \end{eqnarray}
 These results are already known \cite{barger,griffiths,cheng}.

Now, if we assume that $A^\mu$ and $Z^\mu$ transform as $Z_\mu\rightarrow Z_\mu+\lambda_z \partial_\mu \theta$ and $A_\mu\rightarrow A_\mu+\lambda_a \partial_\mu \theta$, then using Eq.~\eqref{eq:Z},Eq.~\eqref{eq:A}, Eq.~\eqref{eq:glam1},Eq.~\eqref{eq:lamtf} and Eq.~\eqref{eq:ggp} it can be easily shown that
\begin{equation}
\label{eq:lalz}
\lambda_a=-1 \quad\textrm{and}\quad \lambda_z=0
\end{equation}
Thus we get the known result of QED for $\lambda_a$ being $-1$ and this  value of $\lambda_a$ can be used as a signature of photon in this method.

Now, let us look at FCNC at tree level. As we discussed earlier, it will be left handed in this approach and the interaction terms, contributing to it, are $\overline\psi_{jL}\gamma_\mu\psi_{kL}A_3^\mu$ and $\overline\psi_{jL}\gamma_\mu\psi_{kL}B^\mu$ where $\psi_j$ and $\psi_k$ are fermions of different flavours but with same charge. In spite of writing the FCNC interaction in terms of $A_3^\mu$ and $B^\mu$, we write it in terms of $A^\mu$ and $Z^\mu$ as
\begin{equation}
\mathscr{L}_{FCNC}=-[\overline\psi_{jL}\gamma_\mu\psi_{kL} (h^a_{jkL}A^\mu+h^z_{jkL} Z^\mu)]
\end{equation}
where $q_j=q_k$ but $j\neq k$ and $h^a_{jkL}$, $h^z_{jkL}$ are couplings with $A^\mu$ and $Z^\mu$ respectively. For hermiticity of Lagrangian, $h^a_{kjL}=h^{a*}_{jkL}$ and $h^z_{kjL}=h^{z*}_{jkL}$ where `$*$' symbolizes complex conjugate.

Using Eq.~\eqref{eq:lalz} and demanding local gauge invariance of this Lagrangian, it is easy to show that $h^a_{jk\tau}=0$. But, it is impossible to infer anything about $h^z_{jk\tau}$ as $\lambda_z=0$. So, left handed FCNC at tree level might exist, but it must be mediated by $Z$-boson only.

So, the electroweak Lagrangian will look like
\begin{equation}
\label{eq:lagew}
\begin{split}
&\widetilde{\mathcal{L}}_{EW}=\widetilde{\mathcal{L}}_0-\frac{1}{4} \mathscr{B}^{\mu\nu}\mathscr{B}_{\mu\nu}-\frac{1}{4}\mathscr{F}_a^{\mu\nu}(\mathscr{F}_a)_{\mu\nu}+\overline{\psi}_j(i\slashed{\partial}-m_j)\psi_j\\&-\frac{g}{\sqrt{2}}[ V_{jk}\overline\psi_{jL}\gamma_\mu\psi_{kL} A_+^\mu
+V_{jk}^*\overline\psi_{kL}\gamma_\mu\psi_{jL} A_-^\mu]\delta_{q+q_k,q_j}\\& -h^z_{jk\tau}\overline\psi_{j\tau}\gamma_\mu\psi_{k\tau} Z^\mu
 \delta_{q_j,q_k} -q_j\overline{\psi}_j\gamma_\mu\psi_j A^\mu
\end{split}
\end{equation}
where $a=(1,2,3)$, $\tau=(L,R)$, $\mathscr{B}^{\mu\nu}=\partial^{\mu}B^\nu-\partial^{\nu}B^\mu$ , $(j,k)$ denote different flavours of fermions and $\widetilde{\mathcal{L}}_0$ contains all the kinetic and interaction terms for new gauge bosons. Moreover, $h^z_{jkR}=0$ for $j\neq k$ and  $h^z_{kk\tau}=(g_z/2)(\chi_{k\tau}-\frac{q_k}{q} \sin^2\theta_W)$ with $\chi_{kR}=0$ for all fermions.

 We see that the Lagrangian in Eq.~\eqref{eq:lagew} has no $SU(2)_L$ symmetry, still it can describe electroweak interactions. In order to bring $SU(2)_L$ symmetry in this scenario, one has to demand that \textit{(i)} $V$ matrix is unitary \cite{kobayashi,barger,griffiths,cheng}, \textit{(ii)} $h^z_{jkL}$ is zero for $j\neq k$, \textit{(iii)} $\chi_{jL}=\pm 1/2$  and  \textit{(iv)} masses of all fermions ($m_j$) are zero.  However, the parameters of this model must depend on the energy scale in such a way that at low energy approximation those four conditions hold  true and the Lagrangian seems to have $SU(2)_L$ symmetry.

\section{Masses of gauge bosons} 
\label{mgb}

Now, the mass terms for gauge bosons can be introduced as
\begin{equation}
\mathcal{L}_{mass}=\frac{1}{2}[m_{ph}^2 A^2+ m_1^2 A_1^2+ m_2^2 A_2^2+m_Z^2 Z^2]
\end{equation}
where $m_{ph},$ $m_1,$ $m_2$ and $m_z$ are the masses of photon, $A_1,$ $A_2$ and  $Z$ respectively. If we rewrite this Lagrangian in terms of $A_+^\mu$ and $A_-^\mu$ and demand global gauge invariance of $\mathcal{L}_{mass}$ then we must have $m_1=m_2=m_W \textrm(say)$. Moreover, in order to make $\mathcal{L}_{mass}$ local gauge invariant we need $m_{ph}=0$ as under local gauge transformation $A^\mu\rightarrow A^\mu-\partial^\mu \theta$. But $Z$ boson might be massive as $\lambda_z=0$, Eq.~\eqref{eq:lalz}. With these constrains mass terms for gauge boson can be written in Lagrangian as,
\begin{equation}
\label{eq:lmas}
\widetilde{\mathcal{L}}_{mass}=\frac{1}{2} m_Z^2 Z^2+ m_W^2 A_+^\mu A_{-\mu}
\end{equation}

However, in order to make the theory renormalizable, the masses of gauge bosons must be zero and they should be generated by some other mechanism.

\section{Renormalizability}
\label{ren}

As the values for different parameters of this model are not made fixed, it is not legitimate to ask whether this theory is renormalizable or not; rather one must demand renormalizability to restrict the parameter-space for this model. To clarify the statement, let us consider a concrete example. Let, there is a charge-less scalar field $\varphi$ (not to be confused with Higgs field). From usual field theory, we know that  $\varphi^3$ theory is non-renormalizable whereas $\varphi^4$ theory is renormalizable (in four dimensions)\cite{peskin}. Now, if we have a theory with the interaction term as $\big(\alpha \varphi^3+\beta\varphi^4 \big)$, we cannot not make any statement about renormalizability of the model till $\alpha$ and $\beta$ are made fixed. However, demanding renormalizability, one can constrain the parameter-space for $(\alpha, \beta)$ as $\beta\neq 0$ because for any non-zero $\beta$, the interaction term results in a shifted $\varphi^4$ theory as $\big(\alpha \varphi^3+\beta\varphi^4 \big)=\beta\Big[\big(\varphi_n+\frac{2}{3}\,\varphi_0\big)^4-6\varphi_0^2 \varphi_n^2-\frac{1}{3}\,\varphi_0^4\Big]$ where, $\varphi_0=\frac{\alpha}{4\,\beta}$ and $\varphi_n=\big(\varphi+\frac{\alpha}{12\,\beta}\big)$. Using similar kind of approach, the parameter space for our model can be restricted. 

To ensure renormalizability of the model, we have taken dimension (mass) four interactions only and the masses of gauge bosons are assumed to be zero before symmetry breaking. Though these are necessary conditions for a theory to be renormalizable, they are not sufficient. Actually, it would take a huge effort which is beyond the scope of this paper to find all the conditions on parameters for renormalizability. However, demanding renormalizability cannot ensure the manifestation of $SU(2)_L$ symmetry in the theory. To elaborate, we demonstrate an exemplary renormalizable model (a subset of our model) which can describe electroweak interactions despite not being SM.

Let us consider a model with  \textit{(i)} $V$ matrix to be  unitary, \textit{(ii)} $h^z_{jkL}$ to be zero for $j\neq k$ and \textit{(iii)} $\chi_{jL}$ to be $\pm 1/2$; i.e. it fulfils three out of four conditions (except the fourth one) for $SU(2)_L$ symmetry, as discussed in last paragraph of section \ref{nci}. Then, renormalizability, anomaly cancellation, tree level unitarity occur automatically as we know from QED and QCD that non-zero fermion masses do not affect renormalizability of a theory. Nevertheless, non-zero fermion masses obstruct $SU(2)_L$ symmetry to emerge in this toy-model. So, in principle, there could exist a renormalizable model (different from SM) interpreting electroweak interactions.

\section{Higgs field}

 To generate masses for gauge bosons, we need one real scalar field $\Phi$  (in ``unitary gauge") with the Lagrangian
\begin{equation}
\label{eq:scalar}
\begin{split}
\widetilde{\mathcal{L}}_{scalar}=\frac{1}{2}[\partial_\mu \Phi \partial^\mu\Phi+\kappa \Phi^2-\omega\Phi^4+\frac{g^2}{2} V_{\Phi\Phi}^2A_+^\mu A_{-\mu}\Phi^2+g_z^2\chi_\phi^2 Z^2 \Phi^2]
\end{split}
\end{equation}
where $\kappa$ and $\omega$ are some arbitrary constants. The overall $1/2$ factor comes because $\Phi$ is real. We want the gauge bosons to interact with $\Phi$ by same coupling as with fermions. Looking at Eq.~\eqref{eq:lagew}, we assume that $A_+$ or $A_-$ interacts with $\Phi$ by coupling $gV_{\Phi\Phi} /\sqrt{2}$. So, we took the coefficient of $A_+^\mu A_{-\mu}\Phi^2$ to be $g^2  V_{\Phi\Phi}^2/2$. As $\Phi$ is not charged, its coupling with photon is zero and the coefficient of $Z^2 \Phi^2$ is taken to be $g_z^2\chi_\phi^2$. There is no other kind of dimension four interactions between the gauge bosons and $\Phi$ due to Lorentz gauge condition. Now, we see from Eq.~\eqref{eq:scalar} that $\widetilde{\mathcal{L}}_{scalar}$ has a $Z_2$ symmetry because $\Phi\rightarrow-\Phi$ leaves it invariant. On the other hand, scalar potential for $\Phi$ without any interactions with gauge bosons, $\mathcal{V}(\Phi)=-\frac{1}{2}(\kappa \Phi^2-\omega\Phi^4)$ has two minima at $\Phi_{min}=\pm\sqrt{\kappa/2\omega}=\pm v$ (say). So, by picking any one of them (let us take $+v$) as the vacuum expectation value of $\Phi$, we can spontaneously break the $Z_2$ symmetry of Lagrangian. Now, writing $\Phi=v+H$ (where $H$ is Higgs field) and looking at Eq.~\eqref{eq:lmas}, one can identify masses of $W$ and $Z$ bosons as 
\begin{equation}
\label{eq:wzmas}
m_W=\frac{1}{2}gv V_{\Phi\Phi}\quad \textrm{and}\quad m_Z=g_zv\chi_\phi^{}
\end{equation}
and their interactions with Higgs field become $(\frac{1}{4}V_{\Phi\Phi}^2 g^2A_+^\mu A_{-\mu}+\frac{1}{2}g_z^2\chi_\phi^2 Z^2)(H^2+2vH)$. On the other hand, the Higgs field gets a mass $m_H=\sqrt{\kappa}$ along with the self interaction terms $\big[-(\kappa/v)H^3-(\kappa/4v^2)H^4\big]$. Now, using Eq.~\eqref{eq:gz}, Eq.~\eqref{eq:wzmas} and definition of $\rho$ parameter, we get
 \begin{equation}
 \label{eq:rotheta}
 \rho=\frac{(g_z^2/m_Z^2)}{(g^2/m_W^2)}=\frac{V_{\Phi\Phi}^2}{4\chi_\phi^2}\quad\textrm{and}\quad \sin^2\theta_W=1-\frac{m_W^2}{ m_Z^2 \rho}
 \end{equation}
Taking $V_{\Phi\Phi}=1$ and $\chi_\phi=1/2$ one can match all the SM results.

In general, fermionic interactions like $-\zeta_{jk} \overline{\psi}_{j}\psi_{k}\Phi\delta_{q_j,q_k}$ can be added to $\widetilde{\mathcal{L}}_{scalar}$, given by Eq.~\eqref{eq:scalar}. Then, we have to break  the discrete chiral symmetry ($\Phi\rightarrow-\Phi$, $\psi_j\rightarrow i\gamma_5\psi_j$ for all fermions) of the Lagrangian spontaneously by picking any one of the two values of $\Phi_{min}$ as vacuum expectation value of $\Phi$. So, a fermion-mass matrix ($M_\zeta$) can be formed with elements $v\zeta_{jk}$ and due to hermiticity of Lagrangian, $M_\zeta=M_\zeta^\dagger$. But, there was a diagonal fermion-mass matrix ($M$) before spontaneous symmetry breaking (SSB) with diagonal entries $m_j$. Then, the final mass matrix, $M_f=M+M_\zeta$. As $M_f$ is hermitian, it can be diagonalized by unitary transformation and that basis of fermions can be called ``mass basis after SSB". So, in this approach, masses of fermions are partly generated by SSB. However, if they are needed to be generated fully from SSB, we have to start with $M=0.$ In contrast, $M$ is always zero in SM due to its chiral structure and hence masses of fermions fully come from SSB in SM.

In SM, we get a similar kind of Lagrangian for the scalar field if the  SSB is taken in ``unitary gauge". However, the interaction term between fermion and scalar field looks a little bit different there due to the chiral structure of SM. Moreover, it is shown in Ref. \cite{masson,faddeev} that, for SSB under unitary gauge, transformations of different fields in SM look same as in this model. But, in the case of Ref. \cite{masson,faddeev}, $SU(2)_L$ symmetry remains hidden in the Lagrangian as the authors have started with SM. On the contrary, there is no hidden $SU(2)_L$ symmetry in our approach. However, if someone wants to compare SSB in SM under some different gauge (other than unitary gauge) with SSB in this model, he/she should start with four independent real scalar fields, $\Phi_i$ $(i=1,2,3,4)$, where three of them are massless (i.e. $\langle{\Phi_1}\rangle=\langle{\Phi_2}\rangle=\langle{\Phi_3}\rangle=0$) and the fourth one ($\Phi_4$) behaves exactly like $\Phi$, described above. These massless scalar fields can have all other types of interactions like $\Phi_4$ except the mass term. In this case also, the scalar potential $\mathcal{V}(\{\Phi_i\})=-\frac{1}{2}(\kappa \Phi_4^2-\omega_i\Phi_i^4)$ has two minima, situated at $\Phi_{min}=(0,0,0,\pm\sqrt{\kappa/2\omega_4})$, and we have to break  the discrete chiral symmetry ($\Phi_i\rightarrow-\Phi_i$, $\psi_j\rightarrow i\gamma_5\psi_j$ for all fermions and scalar fields) of the Lagrangian spontaneously by picking any one of the two values of $\Phi_{min}$ as vacuum expectation value of $(\{\Phi_i\})$. The fermionic interaction term can  also be modified as $-\zeta_{ijk} \overline{\psi}_{j}\psi_{k}\Phi_i\delta_{q_j,q_k}$. There is another interesting instance which looks similar to SSB of SM. In this case, we start with four independent scalar fields, $\phi_i$ $(i=1,2,3,4)$, having same mass  and couplings. Then the scalar potential $\mathcal{V}(\{\phi_i\})=-\frac{1}{2}(\kappa\phi_i^2-\omega\phi_i^4)$ has minima at $|\phi_{min}|=\sqrt{\sum\phi_{i,min}^2}=\sqrt{\kappa/2\omega}$. Again, the scalar potential has a $SO(4)$ symmetry and due to homomorphism between $SO(4)$ and $SU(2)\times SU(2)$, we can say that there exists some $SU(2)$ symmetry in this scalar potential. However, due to presence of interactions between fermions and scalar fields, there does not exist any $SO(4)$ as well as $SU(2)$ symmetry in the Lagrangian of scalar field, unlike SM. Still, there exists the discrete chiral symmetry ($\phi_i\rightarrow-\phi_i$, $\psi_j\rightarrow i\gamma_5\psi_j$) in the whole Lagrangian. So, by picking a particular value of $\phi_{min}$, e.g. $(0,0,0,\sqrt{\kappa/2\omega})$, as the vacuum expectation value of $(\{\phi_i\})$ we break this symmetry spontaneously.

\section{Conclusion}
The parameter-space of this model is larger than that of SM. Using data for electroweak precision tests and other experiments involving weak interaction, one can constrain the parameter-space of this theory. But, the input parameters should be chosen very careful. For example, we cannot use the value of $G_F$, obtained from experimentally measured decay rate of muon, as an input parameter. In SM, muon decay at tree level occurs through $W$-mediated charged current interaction only and using the relation $\frac{G_F}{\sqrt{2}}=\frac{g^2}{8m_W^2}$, we get the experimental value of $G_F$. But, in our approach, it is also possible through $Z$-mediated FCNC currents \big(i.e. $\mu^+\rightarrow e^+ Z(\rightarrow \nu_i\overline{\nu}_j)$ where $(i,j)=(e,\mu,\tau)$\big). Still, using the experimental data for muon decay rate, we can get constrain on the FCNC couplings ($h_{jkL}^z$) for this process. However, if the FCNC couplings for this mode is assumed to be very small with respect to $g$, then one can use the value $G_F$ as an input parameter. Again, due to lack of knowledge about $\rho$ parameter, we cannot estimate the value of $\sin^2\theta_W$ using Eq.~\eqref{eq:rotheta} with measured values of $m_Z^{}$ and $m_W^{}$. On the other hand, the value of $\sin^2\theta_W$, extracted from M{\o}ller scattering \cite{moller}, needs $G_F$ as an input parameter. So, one should start from scratch to look for the restrictions coming from electroweak precision tests and other experiments. On the other hand, to maintain a proper quantum field theoretic approach, one should first quantize this model in path integral formalism by introducing appropriate gauge fixing term and Faddeev-Popov ghosts, then renormalize it and at the end look for different quantum corrections coming from various loops.

However, we have shown that a ``generalized $U(1)_Q$ gauge symmetry" can produce electroweak interactions without demanding any $SU(2)_L$ gauge invariance. Interactions among gauge bosons, charged current interactions, neutral current interactions everything comes naturally once generalized $U(1)_Q$ gauge symmetry is demanded. In addition, scopes for non-unitary  CKM matrix and FCNC at tree level are also discussed. These FCNC interactions can contribute to several precesses like muon decay and leptonic tau decay in tree level, processes involving ``penguin diagrams" (e.g. $b\rightarrow sl^+l^-$) in tree level, neutrino oscillations in one loop level and so on. On the other hand, though the current experimental data for $V$-matrix elements are consistent with SM, the inclusion of FCNC vertices might change the values of different elements in matrix $V$  leading to its non-unitarity. Non-unitary $V$ matrix in lepton sector can lead to non-universality of leptons which can be used for describing the discrepancies between SM prediction and experimental results of various observables in different decay modes, e.g. $R_D - R_{D^*}$ \cite{d*1,d*2,d*3} and $R_K-R_{K^*}$ \cite{k*1,k*2,k*3}. Again, Non-unitary $V$ matrix can also incorporate possibilities for fourth generation fermions more easily than SM. Similarly, in future, if some other electroweak gauge bosons \cite{fifth1,fifth2,new-bos1,new-bos2,new-bos3,new-bos4,fcnc1,fcnc2,fcnc3,fcnc4}  are  discovered experimentally, they can be handled very comfortably in this approach without assuming any new symmetry of Lagrangian. In a nutshell, this approach shows that  demanding $SU(2)_L\times U(1)_Y$ invariance of Lagrangian to describe electroweak interactions is a ``sufficient" but ''not necessary" condition. This model provides more general description of electroweak interactions and SM can be considered as a special case of it. Use of this approach in other gauge theories may lead to some interesting inferences.

\section*{Acknowledgements}
I thank Rahul Sinha and G. Rajasekaran for their valuable comments and suggestions. I also thank N. Mohammedi, Dibyakrupa Sahoo and Arnab Rudra for some extremely useful comments and discussions.

\end{document}